\documentclass[10pt]{article}

\usepackage{amsmath,graphicx}

\def\be{\begin{equation}}
\def\ee{\end{equation}}

\begin{document}
\begin{center}
{\large{\bf On the first $G_1$ stiff fluid spike solution in General Relativity}}
\end{center}

\begin{center}
{\Large\bf }
\vspace{.3in}
{\bf A A Coley, D Gregoris}, 
\\Department of Mathematics and Statistics, Dalhousie University,
\\Halifax, Nova Scotia, Canada B3H 3J5
\\Email: aac@mathstat.dal.ca ,  danielegregoris@libero.it
\vspace{.1in}
\\ {\bf W C Lim},
\\Department of Mathematics and statistics, University of Waikato,
\\ Private Bag 3105, Hamilton 3240, New Zealand
\\Email: wclim@waikato.ac.nz
\vspace{.1in}
\vspace{0.2in}

\end{center}

\begin{abstract} Using the Geroch transformation we obtain the first example of an exact stiff fluid spike solution 
to the Einstein field equations in a closed form exhibiting a spacelike $G_1$ 
group of symmetries (i.e., with a single isometry).  This new solution is of Petrov type I and exhibits a spike
crossing which persists to the past, which allows us to better understand spike crossings in the context
of structure formation.  \end{abstract}


\section{Introduction}
The search for new exact solutions to the Einstein field equations (EFE) and their physical interpretation have been a fascinating pursuit since the first days of General Relativity. 
Due to the nonlinearity of the EFE it is often necessary to resort to assuming some symmetries on the solution. 
Most of the solutions that have been derived so far are discussed and classified in \cite{book:Exactsol2002}.
In this paper we derive a new analytic solution of the EFE admitting a $G_1$ (i.e., a one dimensional)
group of motions with a non-null Killing vector field (KVF)
and whose matter source is a perfect fluid obeying the ideal equation of state characterizing stiff matter.%
\footnote{Note that analytical $G_1$ solutions in vacuum are discussed in page 271 of \cite{book:Exactsol2002}, 
while exact $G_1$ solutions with a null KVF are summarized in page 379 of the same book. 
Senovilla and Sopuerta~\cite{art:SenovillaSopuerta1994} found some Petrov type D $G_1$ solutions with a non-null KVF.
These were generalized by Koutras and Mars~\cite{art:KoutrasMars1997}, and are also of Petrov type D. See also~\cite[pages 504, 569]{book:Exactsol2002}.}

We will use the Geroch transformation to generate the new solution \cite{art:Geroch1971, art:Geroch1972, art:GarfinkleGlassKrisch1997}, 
show that it admits only one KVF, and demonstrate its spike crossing properties.
Since the spike is persisting close to the initial singularity our solution constitutes an explicit example 
in which the Belinskii, Khalatnikov and Lifshitz (BKL) locality conjecture~\cite{art:LK63, art:BKL1970, art:BKl1982} does not apply~\cite{art:Limetal2006}.

This paper is organized as follows: after a review about the solution-generating Geroch transformation, we apply it to two seed solutions with stiff fluid
-- the hyperbolic Friedmann-Lemaitre (FL) solution and a spatially homogeneous but anisotropic Bianchi type V solution.
We will then compare these two solutions in the late-time regime. 
We then analyze the mathematical and physical properties of the latter solution. 
We conclude with a brief discussion.
We will use the signature $(-, +, +, +)$ and geometric units.

\section{The Geroch transformation: an overview}

Let $g_{ab}$ denote a known exact solution to the EFE admitting a KVF $\xi_a$ orthogonal to the fluid four velocity; 
assume also that the matter source is given by a perfect fluid whose energy density and pressure are related by the equation of state $p=\rho$. 
Introduce the norm of the KVF $\lambda:= \xi_a \xi^a$ and the twist $\omega_a:=\epsilon_{abcd} \xi^b \nabla^c \xi^d$, 
$\epsilon_{abcd}$ being the completely antisymmetric Levi-Civita symbol. 
Then we can integrate $\omega_a$ to find a scalar $\omega$ with the property $\omega_a = \nabla_a \omega$, 
and form $\alpha_a$ and $\beta_b$ solutions to the following system of partial derivative equations:
\begin{eqnarray}
\nabla_{[a} \alpha_{b]}\,&=&\, \frac12 \epsilon_{abcd}  \nabla^c \xi^d \\
\xi^a \alpha_a \,&=&\, \omega \\
\label{diffbeta}
\nabla_{[a} \beta_{b]}\,&=&\, 2\lambda \nabla_a \xi_b +\omega \epsilon_{abcd}  \nabla^c \xi^d \\
\label{norma}
\xi^a \beta_a\,&=&\, \omega^2 + \lambda^2 -1 \,.
\end{eqnarray}
We can now introduce a rotation parameter $\theta$ in terms of which we have the two new quantities $\tilde \lambda$ and $\eta_a$:
\begin{eqnarray}
\frac{\lambda}{\tilde \lambda} \,&:=&\, \cos^2 \theta + (\omega^2 +\lambda^2) \sin^2 \theta - 2 \omega \sin \theta \cos\theta \\
\eta_a \,&:=&\, \frac{\xi_a}{\tilde \lambda}+2 \cos\theta \sin\theta \alpha_a - \sin^2\theta \beta_a \,.
\end{eqnarray}
Finally, the new metric can be written in terms of the above quantities as:
\be
\label{new1}
{\tilde g}_{ab}\,=\, \frac{\lambda}{\tilde \lambda}\left(g_{ab}-\frac{\xi_a \xi_b}{ \lambda} \right)+{\tilde \lambda}\eta_a \eta_b\,.
\ee
For simplicity, we set $\theta= \frac{\pi}{2}$, which leads to the decoupling of $\alpha_a$ from the solution. As a result, we have the following simpler equations
\begin{eqnarray}
\frac{\lambda}{\tilde \lambda} \,&:=&\,   (\omega^2 +\lambda^2)  \\
\eta_a \,&:=&\, \frac{\xi_a}{\tilde \lambda} -  \beta_a \,.
\end{eqnarray}
The new metric in this case is given by
\be
{\tilde g}_{ab}\,=\, (\omega^2 + \lambda^2)g_{ab} + \frac{\lambda}{\omega^2 + \lambda^2}\beta_a \beta_b - 2\xi_{(a} \beta_{b)}\,,
\ee
where round parentheses denote symmetrization. 

The new solution admits the KVF $\xi_a$ and satisfies the equation of state of an ideal stiff fluid whose energy density is given by:
\be
\rho  \to {\tilde \rho}=\frac{\rho}{\lambda^2 + \omega^2}\,.
\ee

\section{Applying the Geroch transformation to the FL seed}

In this section we will derive the Geroch transform for a hyperbolic FL seed whose metric is given by \cite{book:Exactsol2002}
\be
\label{seedF}
ds^2=A^2(t)[-dt^2+dx^2+e^{2x}(dy^2+dz^2)]\,,
\ee
assuming that the comoving matter content is described by the stress-energy tensor $T^{a}{}_{b}=[-\rho_F(t), \rho_F(t), \rho_F(t), \rho_F(t)]$. Solving the EFE $G_{ab}=T_{ab}$
we get explicitly\footnote{Note that we are interested in a {\it particular} solution.}:
\be
ds^2=\sinh (2t)[-dt^2+dx^2+e^{2x}(dy^2+dz^2)]\,
\ee
and
\be
\rho_F=\frac{3}{\sinh^3 (2t)}\,.
\ee
Then as a KVF we select $\xi=\partial_x-y\partial_y-z\partial_z$ to match the choice we make in the next section.

For generating the transformed metric (\ref{new1}) we start by applying the coordinate transformation 
\be
\label{coordtra}
t=T\,, \qquad x=X\,, \qquad y=Ye^{-X}\,, \qquad z=Ze^{-X}
\ee
to the seed (\ref{seedF}) to obtain a ``rotated" metric:
\begin{eqnarray}
\label{rotatedF}
&& g_{11F}=-\sinh(2T)\,, \qquad g_{22F}=\sinh(2T)(Y^2+Z^2+1)\nonumber\\
&& g_{23F}=-Y\sinh(2T)\,, \qquad g_{24F}=-Z\sinh(2T)\nonumber\\
&& g_{33F}=\sinh(2T)\,, \qquad g_{44F}=\sinh(2T)
\end{eqnarray}
and a ``rotated" KVF 
\be
\label{killing}
\xi=\partial_X\,.
\ee
The norm of (\ref{killing}) with respect to the metric (\ref{rotatedF}) is given by 
\be
\lambda_F=\sinh(2T)(Y^2+Z^2+1)\,,
\ee
while its twist is given by
\be
\omega_F=\omega_{0F}
\ee
where $\omega_{0F}$ denotes a constant that we will set to zero for simplicity in what follows. 
Then we must explicitly solve the system of partial differential equations (\ref{diffbeta}) for computing $\beta_a$. 
Taking into account again that we are looking only for a particular solution we can make the following simplifications:
\begin{itemize}
  \item $\beta_a$ has vanishing temporal component
  \item $\beta_a$ does not depend on the coordinate $X$ since the KVF $\partial_X$ is preserved by the Geroch transformation
  \item $\beta_X (T,Y,Z) =\sinh^2(2T)(Y^2+Z^2+1)^2 - 1$ where we used the normalization condition (\ref{norma}).
\end{itemize}
Therefore, the non trivial partial differential equations (\ref{diffbeta}) to be solved are
\begin{eqnarray}
\frac{\partial \beta_Y(T,Y,Z)}{\partial_T}&\,=\,& -2Y[Y^2+Z^2+1]\sinh(4T) \\
\frac{\partial \beta_Z(T,Y,Z)}{\partial_T}&\,=\,& -2Z[Z^2+Y^2+1]\sinh(4T) \\ 
\frac{\partial \beta_Z(T,Y,Z)}{\partial_Y}&\,=\,&\frac{\beta_Y(T,Y,Z)}{\partial_Z}\,.
\end{eqnarray}
An explicit solution to this system of equations is:
\be
\beta_Y=-\frac12 Y \cosh(4T)[Y^2+Z^2+1]\,, \qquad \beta_Z=-\frac12 Z \cosh(4T)[Y^2+Z^2+1]\,,
\ee
where we have chosen trivial integration constants when integrating.

Substituting all of the considerations we have presented in this section into (\ref{new1}) we finally get the Geroch transformed metric for (\ref{seedF}):
\begin{eqnarray}
\label{transformedF}
&& {\tilde g}_{11F}=-\sinh^3(2T)[Y^2+Z^2+1]^2\,, \qquad {\tilde g}_{22F}=\frac{1}{\sinh(2T)(Y^2+Z^2+1)} \nonumber\\
&& {\tilde g}_{23F}=\frac{Y}{2\sinh(2T)} \,, \qquad {\tilde g}_{24F}=\frac{Z}{2\sinh(2T)}  \nonumber\\
&& {\tilde g}_{33F}=\frac{(Y^2+Z^2+1)[4(Z^2+1)\sinh^4(2T) +Y^2]}{4\sinh(2T)}  \nonumber\\
&& {\tilde g}_{34F}=\frac{\cosh(4T) (2-\cosh(4T))ZY(Y^2+Z^2+1)}{4\sinh(2T)}  \nonumber\\
&& {\tilde g}_{44F}=\frac{(Y^2+Z^2+1)[4(Y^2+1)\sinh^4(2T) +Z^2]}{4\sinh(2T)} \,. \nonumber\\
\end{eqnarray}
Notice that this metric is symmetric under $ Y \leftrightarrow Z$ and is an exact solution to the EFE for stiff matter whose energy density is given by 
\be
\label{rhoF}
{\tilde \rho}_F\,=\, \frac{3}{\sinh^5 (2T)[Y^2+Z^2+1]^2}\,.
\ee

\section{Applying the Geroch transformation to the Bianchi type V seed}

In this section we will derive the Geroch transform for a particular spatially homogeneous but anisotropic Bianchi type V 
(BV in short) seed whose metric is given 
by\footnote{Note that we are using the same system of coordinates as in the previous section.} \cite{book:Exactsol2002}
\be
\label{seedB}
ds^2=A^2(t)[-dt^2+dx^2]+B^2(t)e^{2x}dy^2+C^2(t)e^{2x}dz^2\,,
\ee
assuming that the comoving matter content is described by the stress-energy 
tensor $T^{a}{}_{b}=[-\rho_B(t), \rho_B(t), \rho_B(t), \rho_B(t)]$. Solving the EFE $G_{ab}=T_{ab}$
we get explicitly the following particular solution \cite{art:Ruban1977a, art:Ruban1977b}:
\be
\label{seedB2}
ds^2=\sinh (2t)[-dt^2+dx^2+e^{2x}(\tanh t dy^2+(\tanh t)^{-1} dz^2)]\,
\ee
and
\be
\rho_B=\frac{1}{4\sinh^3 (t)\cosh^3 (t)}=\frac23 \rho_F\,.
\ee
There are three KVFs to choose from \cite{book:Exactsol2002}: 
\be
\xi=\partial_y\, \qquad \xi=\partial_z\, \qquad \xi=\partial_x-y\partial_y-z\partial_z\,.
\ee
We shall choose the general case, which is a linear combination of all three. But a simple translation in $y$ and $z$ coordinates can be used to absorb the first two KVFs.
Thus the general case is equivalent to choosing the third KVF.

For generating the transformed metric (\ref{new1}) we start by applying the coordinate transformation (\ref{coordtra})
to the seed (\ref{seedB2}) obtaining a ``rotated" metric:
\begin{eqnarray}
\label{rotatedB}
&& g_{11B}=-\sinh(2T)=g_{11F}\,, \qquad g_{22B}=\frac{\sinh(2T)(\tanh^2(T) Y^2 + Z^2 +\tanh T)}{\tanh T} \nonumber\\
&&   g_{23B}=-Y\sinh(2T)\tanh T =\tanh T g_{23F}\,, \qquad g_{24B}=-\frac{Z \sinh (2T)}{\tanh T}=\frac{ g_{24F}}{\tanh T}\nonumber\\ 
&& g_{33B}=\sinh(2T)\tanh T=\tanh T g_{33F} \,, \qquad g_{44B}=\frac{ \sinh (2T)}{\tanh T}=\frac{ g_{44F}}{\tanh T}\,,
\end{eqnarray}
where we write the metric in this way to make explicit that it reduces to the rotated FL seed (\ref{rotatedF}) in the limit $\tanh T \to 1$. 
The ``rotated" KVF is given by (\ref{killing}), whose norm with respect to the metric (\ref{rotatedB}) is given by 
\be
\label{lambdaB}
\lambda_B=\frac {\sinh(2T)(\tanh^2(T) Y^2+Z^2+\tanh T)}{\tanh T}\,,
\ee
which reduces to $\lambda_F$ in the limit $\tanh T \to 1$,
while its twist depends on the spatial coordinates:
\be
\label{omegaB}
\omega_B=2YZ + \omega_{0B},
\ee
where $\omega_{0B}$ denotes a constant that we will set to zero for simplicity in what follows. 
Then we must explicitly solve the system of partial differential equations (\ref{diffbeta}) for computing $\beta_a$ specifically for this seed. 
We can, however, apply all the same simplifications we discussed in the previous section, just changing the normalization condition to
\be
\beta_X (T, Y, Z)= \frac {\sinh^2(2T)(\tanh^2(T) Y^2+Z^2+\tanh T)^2}{\tanh^2 T}-1+4 Y^2 Z^2\,.
\ee
In this case the non trivial partial differential equations (\ref{diffbeta}) to be solved reduce to
\begin{eqnarray}
\frac{\partial \beta_Y(T,Y,Z)}{\partial_T}&\,=\,& -4Y \sinh(2T)[\sinh(2T)+2(Y^2+Z^2)\sinh^2 T] \\
\frac{\partial \beta_Z(T,Y,Z)}{\partial_T}&\,=\,& -4Z \sinh(2T)[\sinh(2T)+2(Y^2+Z^2)\cosh^2 T] \\ 
\frac{\partial \beta_Z(T,Y,Z)}{\partial_Y}&\,=\,&\frac{\beta_Y(T,Y,Z)}{\partial_Z}-16YZ \cosh^2 T +8YZ\,.
\end{eqnarray}
An explicit solution to this system of equations is
\begin{eqnarray}
\label{betab}
\beta_Y&=&-\frac12 Y \sinh(4T) -4 \sinh^4 T Y(Y^2+Z^2) +2YT \\
 \beta_Z&=&-\frac12 Z \sinh(4T) -4 \cosh^4 T Z(Y^2+Z^2) +2ZT             \,, \nonumber
\end{eqnarray}
where once again we have chosen trivial integration constants.

Substituting all the considerations we have presented in this section into (\ref{new1}) we finally get the Geroch transformed metric for (\ref{seedB}). 
We stress the fact that this is an exact solution to the EFE for stiff matter whose energy density is given by 
\be
\label{rhoB}
{\tilde \rho}_B\,=\, \frac{1}{4\sinh^3 T\cosh^3 T \left[\frac {\sinh^2(2T)(\tanh^2(T) Y^2+Z^2+\tanh T)^2}{\tanh^2 T}    +4Y^2Z^2\right]}\,,
\ee
and that remarkably the new solution is given in a closed form in terms of elementary functions.

\section{Late time behavior of the new solution}

In this section we will analyze the late time properties of the two metrics we have generated through the Geroch
transformation.  We know that the FL seed is just a particular subclass of the BV seed and that both spacetimes tends to
the Milne spacetime at late times; we now want to understand if this is also the case for the transformed metrics.  We
will not consider the metric tensors themselves, but a set of scalar curvature invariants that can be derived from them
and some related appropriately defined Hubble-normalized scalars.  

The scalar curvature invariants we consider are  the Ricci scalar ${^1}I \equiv R$  
(which is equivalent, by taking the trace of the EFE, to minus 
the energy density $\rho$), 
${^2}I \equiv R^{-4} R_{,a}R^{,a}$
(equivalent to the norm of the gradient of
the inverse energy density: $|| \nabla_a (1/\rho)||$), and,
iterating the process, the norm of the gradient of the second invariant,  
${^3}I \equiv \nabla_a {^2}I\nabla^a {^2}I$.

First of all the energy densities of the two metrics compare as
\be
{\tilde \rho}_B=\frac23 \frac{\lambda^2_F}{\lambda^2_B + \omega^2_B} {\tilde \rho}_F \,,
\ee
which at late times reads as 
\be
   \frac{{^1I}_B}{{^1I}_F} \equiv \frac{{\tilde \rho}_B}{{\tilde \rho}_F}=\frac23 +O(e^{-2T})\,.
\ee
We also have that:
\be    
    \frac{^2I_B}{^2I_F}=\frac32+O(e^{-2T})\,, \qquad \frac{^3I_B}{^3I_F}=\frac{81}{16}+O(e^{-2T})\,.
\ee
These results show that at late times the ratio of the invariants we have considered for the two metrics converge 
to constant numerical values.

We will now study the late-time behaviour of the Hubble-normalized variables. We begin by introducing the 
unit timelike comoving observer four-velocity
\be
u^a=\frac{dx^a}{d\tau}\,, \qquad u_a u^a=-1\,,
\ee
where $\tau$ denotes the proper time long the observer worldline. We can then introduce the projector orthogonal to this four velocity:
\be
h_{ab}=g_{ab}+u_a u_b\,, \qquad h_{ab}u^b=0\,
\ee
the spacelike symmetric trace-free shear tensor $\sigma_{ab}={\tilde \nabla}_{ \langle a}u_{b \rangle}$ and the rate of volume  expansion $\Theta={\tilde \nabla}_a u^a$, where the latter gives the Hubble function as $\Theta=3H$, where we made use of 
angle brackets to denote  the orthogonally projected
symmetric trace-free part of  a rank 2 tensor:
\be
T^{\langle ab \rangle}=[h^{(a}{}_{c}h^{b)}{}_{d}-\frac13 h^{ab}h_{cd}]T^{cd}
\ee
and where the fully orthogonally projected covariant derivative reads as
\be
{\tilde \nabla}_c T^{a}{}_{b}= h^{a}{}_{d} h^{e}{}_{b} h^{f}{}_{c}     \nabla_f T^{d}{}_{e}\,.
\ee
We can now introduce the matter parameter, the curvature parameter and the Hubble-normalized shear (relative to the timelike
4-velocity):
\be
\Omega_m=\frac{\rho}{3H^2}\,, \qquad \Omega_k=-\frac{^3\! R}{6H^2}\,, \qquad \Sigma^2=\frac{\sigma^2}{3H^2}=\frac{\sigma_{ab} \sigma^{ab}}{6H^2}\,,
\ee
where $^3\! R$ is the Ricci scalar of the 3-dimensional space whose metric is $h_{ab}$.
As observer four-velocity we choose
the comoving matter velocity
\begin{eqnarray}
u^a&=&\frac{1}{\sinh^{\frac32}(2T)\, (Y^2+Z^2+1)}\delta^a_T \\
 u^a&=&\frac{1}{2\sqrt{\sinh(2T) {\mathcal A}}}\delta^a_T\\
{\mathcal A}&=& [(Y^2+Z^2)^2+1]\cosh^4 T+[ \sinh(2T) (Y^2+Z^2)-2Y^4-2Y^2Z^2-1]\cosh^2 T \nonumber\\
&-&\sinh(2T) Y^2 +Y^2(Y^2+Z^2)\, \nonumber
\end{eqnarray}
for the transformed FL and BV seeds, respectively.

The late time limit for the transformed FL seed shows that
\begin{gather}
H=\frac{10 \sqrt{2}}{3(Y^2+Z^2+1)}e^{-3T}+O(e^{-7T}), \quad \Omega_m=\frac{36}{25}e^{-4T}+O(e^{-8T}),\nonumber\\
 \Omega_k= \frac{9}{25}+O(e^{-4T}), \quad \Sigma^2\equiv \frac{16}{25},
\end{gather}
while at late time for the transformed BV metric we get 
\begin{gather}
H=\frac{10 \sqrt{2}}{3 (Y^2+Z^2+1)}e^{-3T}+O(e^{-5T}), \quad \Omega_m=\frac{24}{25}e^{-4T}+O(e^{-6T}),\nonumber\\ 
\Omega_k= \frac{9}{25} + O(e^{-2T}), \quad \Sigma^2=\frac{16}{25} + O(e^{-2T}).
\end{gather}

We note that at late times both transformed solutions (and, in addition,  the transformed Milne solution) 
anisotropically approach the same 
vacuum state (as indicated by the various invariants). Indeed, (i)
the transformed solutions do not isotopize since $\Sigma^2$ asymptotes to a non-zero constant value at late times, (ii)
the solutions approach the same vacuum solution at late times anisotropically (in that the comoving matter is shearing),
and (iii)
the asymptotic late time vacuum state is not Milne since its Weyl tensor is non-zero. 

We also checked explicitly that the Gauss contraint
\be
^3\! R=2\rho - \frac23 \Theta^2 +2\sigma^2
\ee
is satisfied during all the time evolution.

\section{Physical and mathematical properties of the new solution}

In this section we will discuss in more detail some of the physical and mathematical properties of the new 
exact solution of the
EFE we derived in the fourth section of this manuscript.

To look for a spiky behavior of the new metric we must consider the inhomogeneity parameter 
\be
\frac{\omega^2 - \lambda^2}{\omega^2 + \lambda^2}\,,
\ee
where $\lambda$ and $\omega$ are given by equations (\ref{lambdaB}) and (\ref{omegaB}) above.
Usually a spike occurs at the place where $\omega=0$ as $\lambda$ becomes small. 
In this case spikes form along the planes $Y=0$ and $Z=0$, with a spike intersection on the line $Y=Z=0$.%
\footnote{If we kept non-zero $\omega_{0B}$ in (\ref{omegaB}) then spikes form along two hyperbolic cylinders $YZ=-\frac{\omega_{0B}}{2}$, and do not intersect.}
As $T \rightarrow 0$ towards the singularity,
\be
\frac{\omega^2 - \lambda^2}{\omega^2 + \lambda^2} \rightarrow \frac{Y^2-Z^2}{Y^2+Z^2},
\ee
which is discontinuous on the intersection $Y=Z=0$. This shows that the spikes persist at the singularity.
We can also try to find the time at which $\frac{\omega}{\lambda}$ reaches a maximum (which yields the narrowest spikes).
Solving $\partial_T(\frac{\omega}{\lambda})=0$, we obtain
\be
T_{max}=\frac14 \ln \frac{Y^2 + Z^2 -1}{Y^2 + Z^2 +1 }\,,
\ee
which is out of range. Thus the spikes become narrowest at the singularity in this case.

\begin{figure}
\begin{center}
    $
    \begin{array}{cc}
  {\includegraphics[scale=0.3, angle=270]{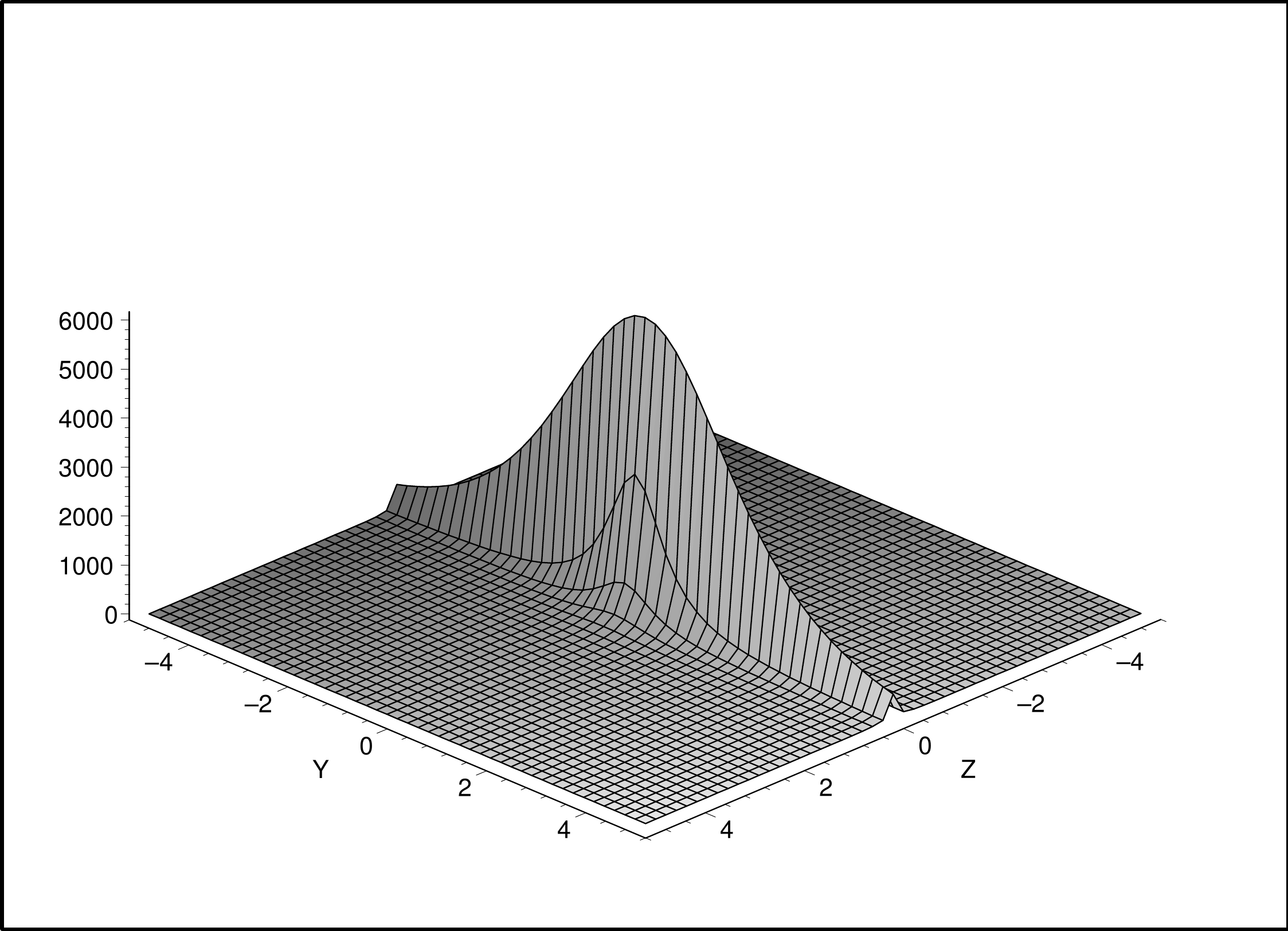}}  &
  {\includegraphics[scale=0.3, angle=270]{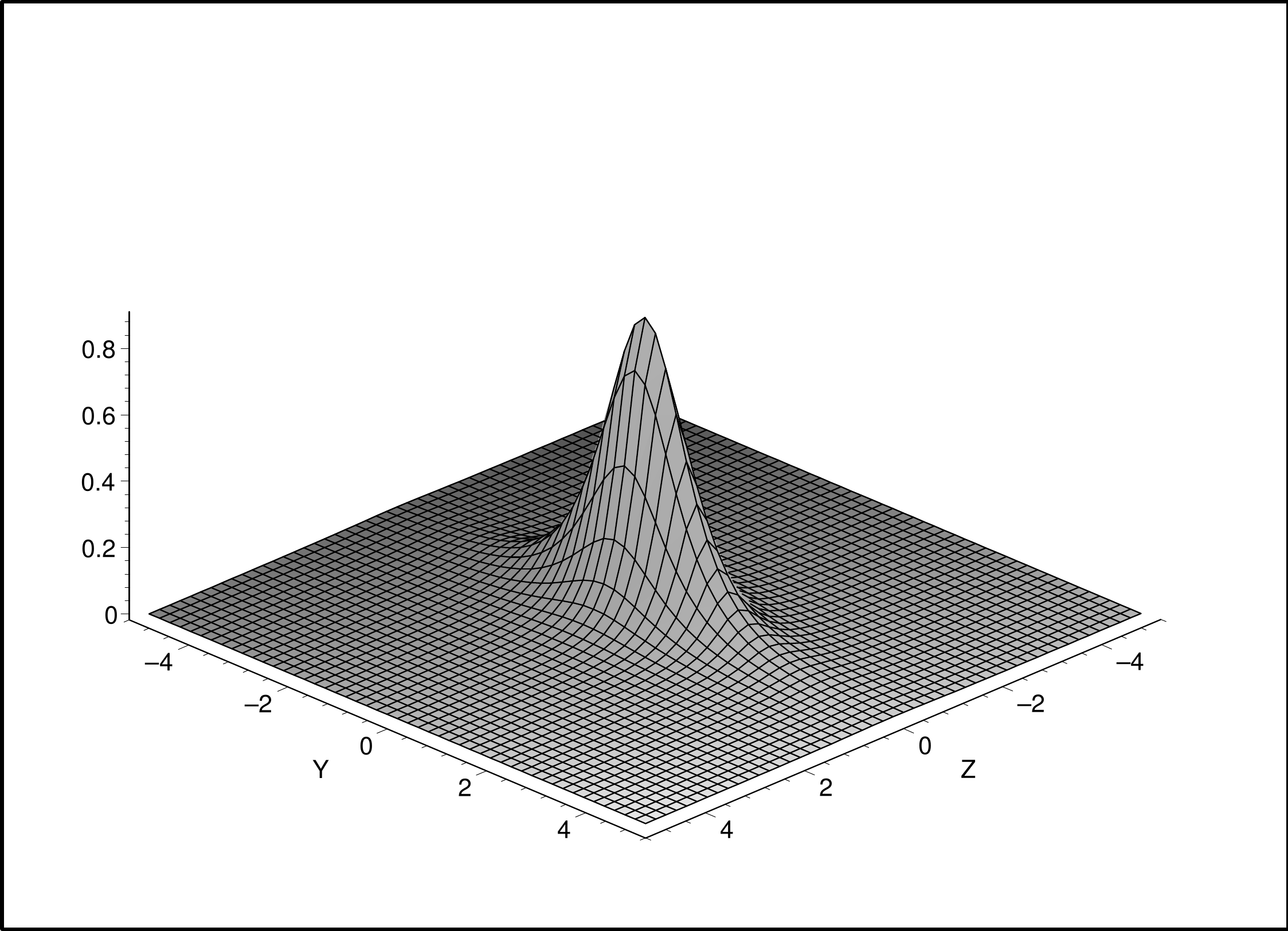}} \cr 
 (a) & (b) \cr
  {\includegraphics[ scale=0.3, angle=270]{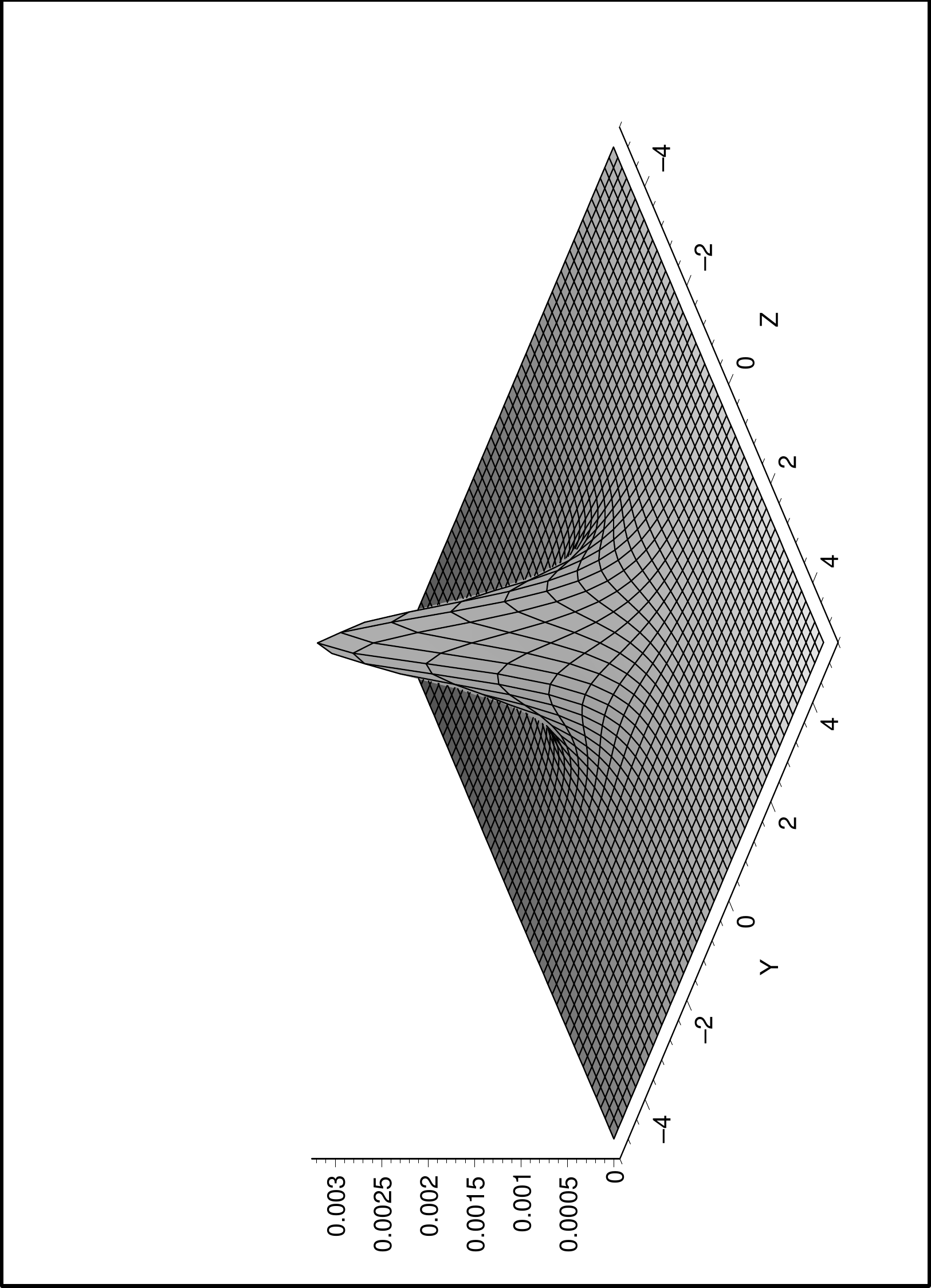}}  &
  {\includegraphics[ scale=0.3, angle=270]{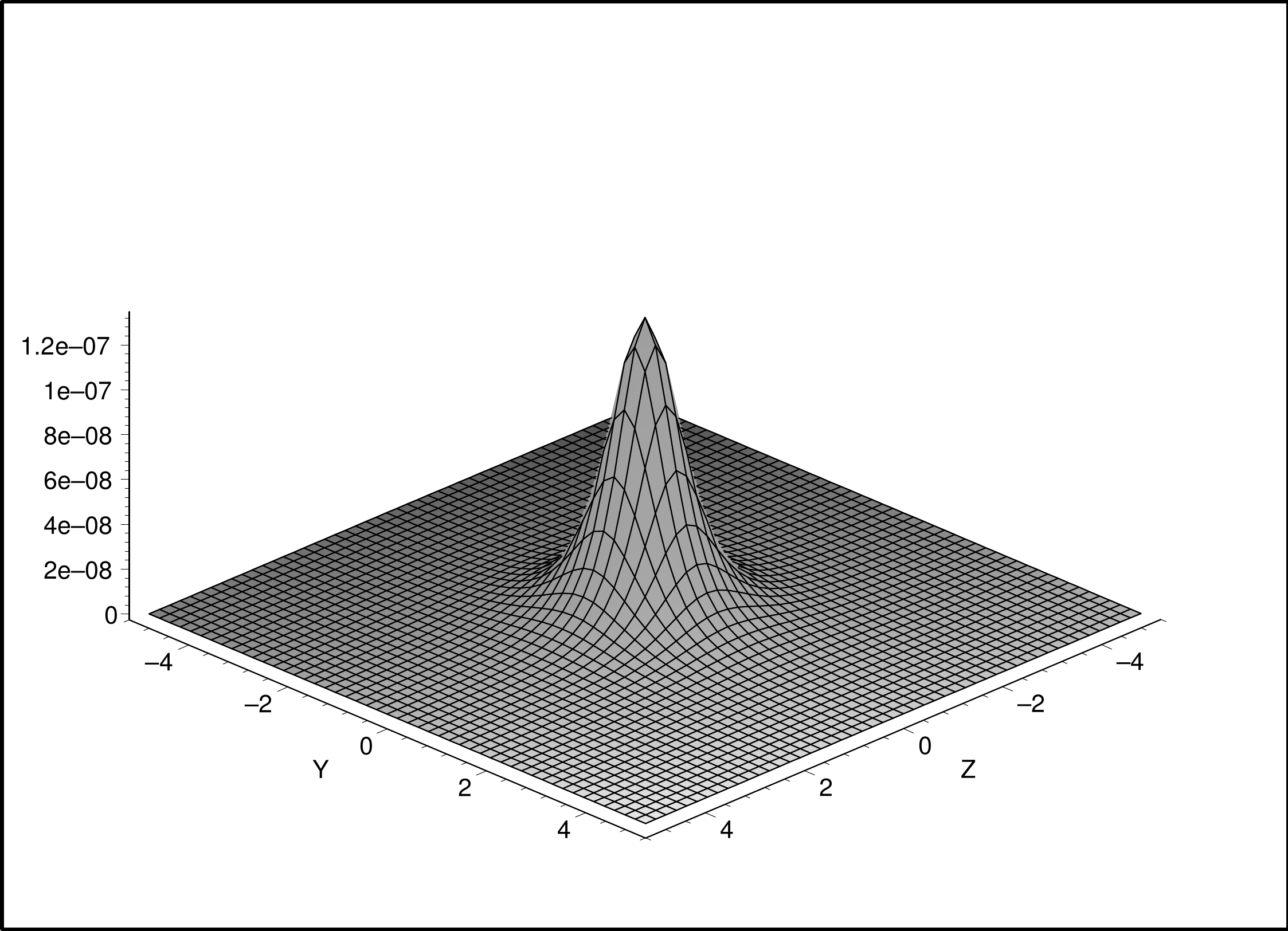}} \cr 
(c) & (d)  \cr
 {\includegraphics[ scale=0.3, angle=270]{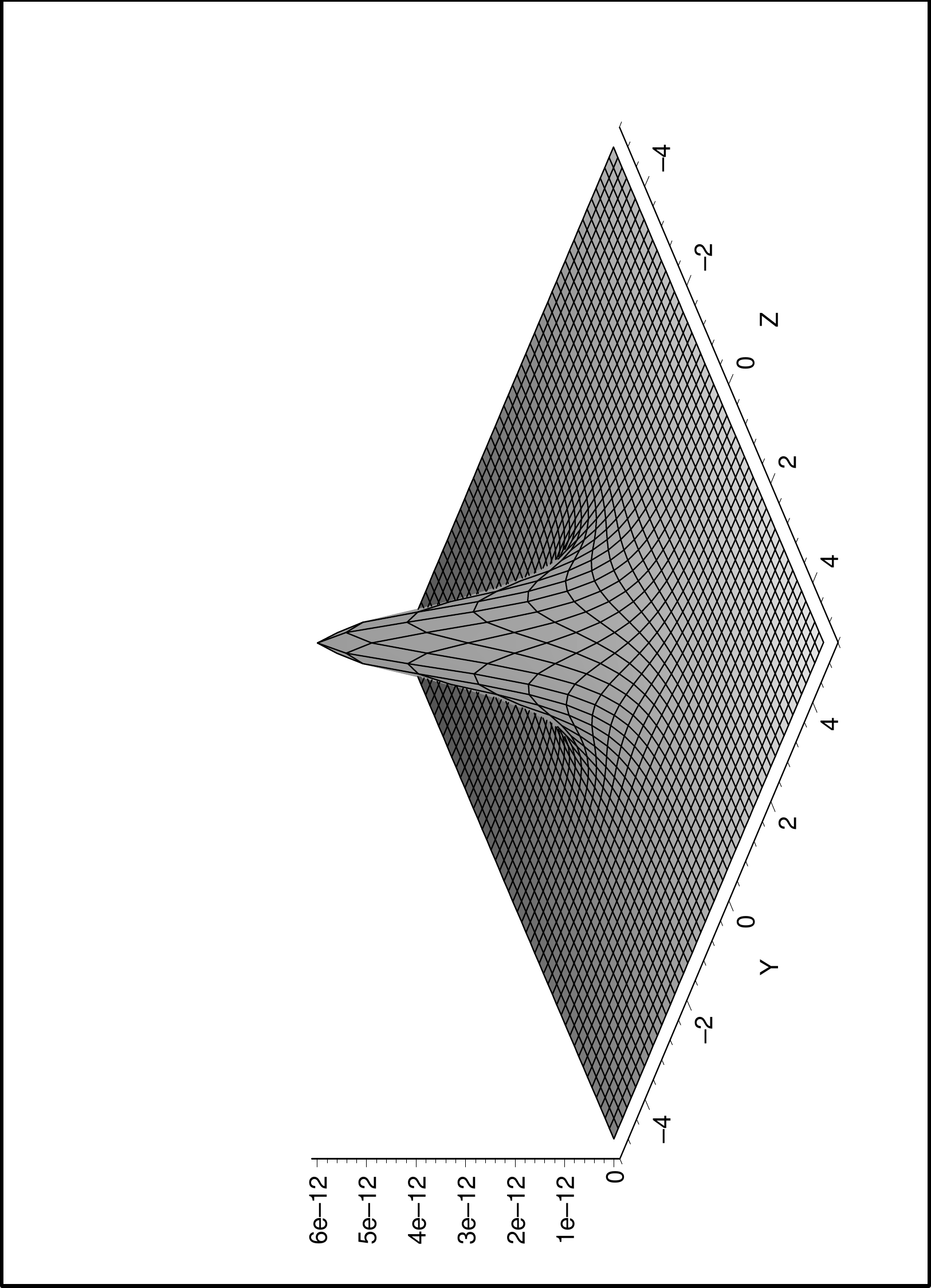}}  &
  {\includegraphics[ scale=0.3, angle=270]{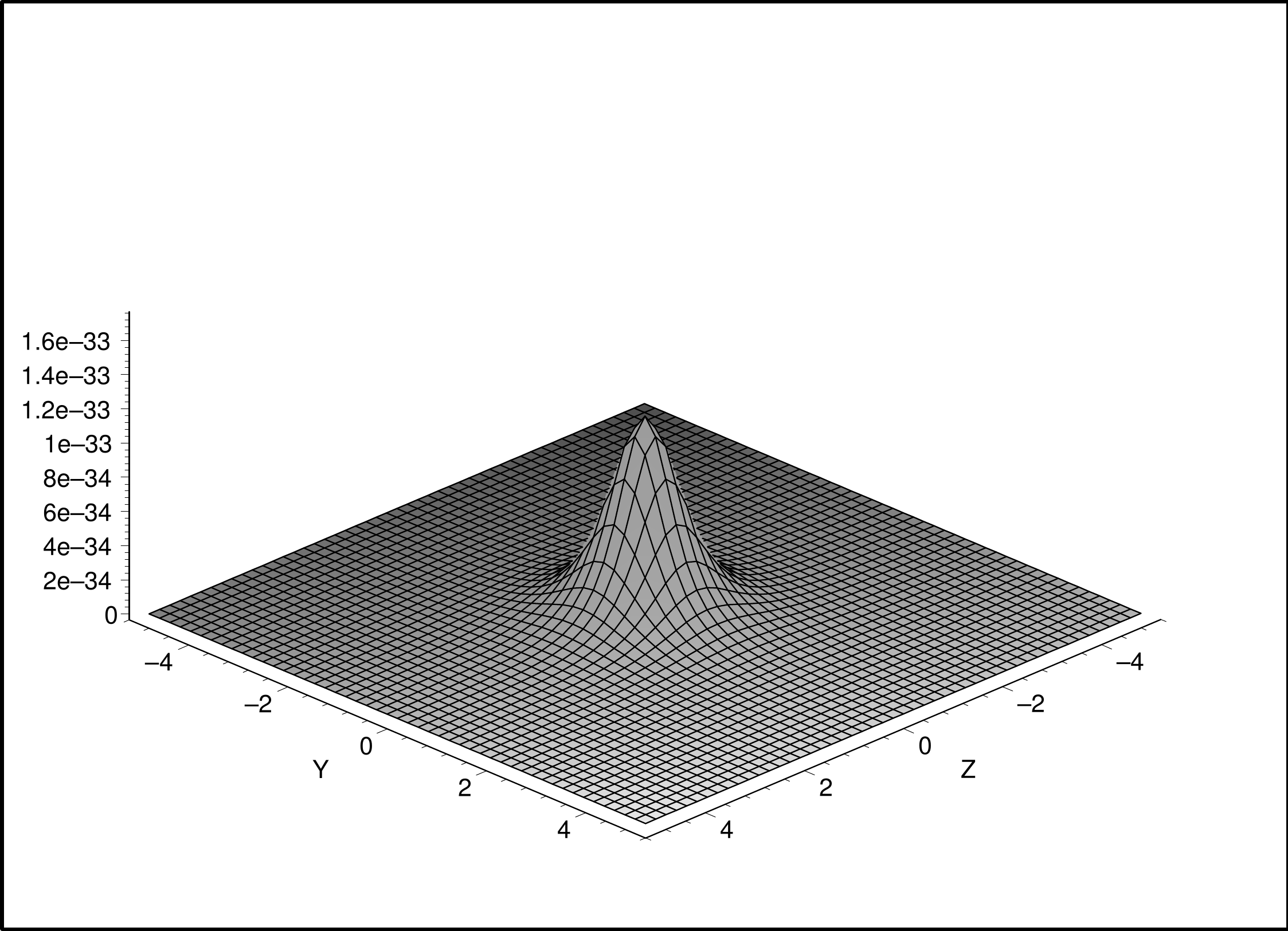}} \cr 
(e)  & (f)  \cr
    \end{array}$
   \end{center}
\caption{We plot the snapshots of the energy density (\ref{rhoB}) for the transformed BV metric in a two-dimensional plane for $Y=-5...5$ and $Z=-5...5$. 
         The time is $T=0.1$, $T=0.5$, $T=1$, $T=2$, $T=3$, $T=8$ in panel (a), (b), (c), (d), (e) and (f) respectively.}
\label{figspike}
   \end{figure}

For making more evident the existence of a spike we plot the energy density (\ref{rhoB}) for the transformed BV metric in figure~\ref{figspike} in the $Y$-$Z$ plane during the time evolution. 
The figure confirms the existence of two intersecting spikes along $Y=0$ and $Z=0$. 
The snapshot at $T=0.1$ (figure~\ref{figspike}(a)) has a visible spike along $Z=0$. 
In all snapshots, the spike along $Y=0$ is indistinct, while the spike at the intersection $(Y,Z)=(0,0)$ is much more prominent than spikes along either $Y=0$ or $Z=0$.
Moreover the spikes disappear at late times.

Finally, we recall that the Geroch transformation preserves the KVF that was applied, so the new solution has at least one KVF.
To show that the new solution does not admit more KVFs, consider the fact that any scalar curvature invariant $I$ satisfies the condition 
\be
\label{invariant_V}
	I_{,a} V^a = 0
\ee
for any KVF $V^a$.
Since $\partial_X$ is already a KVF, all invariants are independent of the $X$ coordinate, and the $X$ component of $V^a$ drops out of equation (\ref{invariant_V}).
We need three invariants such that the system of equations (\ref{invariant_V}) yields trivial solutions for the $T$, $Y$ and $Z$ components of $V^a$.
We utilize the three invariants ${^1}I$,  ${^2}I$, and
${^3}I$ defined earlier.
We then compute the gradient of each invariant and put them into their Jacobian matrix $J$, omitting the vanishing $X$-derivative component.
The system of three equations in matrix form is then $J V^a = 0$, omitting the $X$ component.
We then evaluate the Jacobian matrix at an arbitrary point, say $(T,Y,Z) = (0.1, 0.2, 0.2)$, and compute the determinant of the matrix.
The determinant is non-zero, and so the system yields trivial solution for the $T$, $Y$ and $Z$ components of $V^a$.
Thus, any KVF of the new solution must have vanishing $T$, $Y$ and $Z$ components. The only such KVF is the KVF $\partial_X$. 
Therefore the new solution is genuinely $G_1$. 
In contrast to previous Petrov type D $G_1$ solutions with non-null KVF, the new solution presented here is of the most general Petrov type I.
In order to demonstrate this we simply note that the asymptotic form of the early-time attractor, which is a Jacobs stiff fluid
solution, is of Petrov type I, and hence the new solution must also be of Petrov type I.

\section{Conclusion}

In this paper we have derived a new $G_1$ stiff fluid solution, described in terms of elementary
functions.
It is of mathematical interest since it is the first explicit solution to show interesting spike crossings,
which completes and complements the physical discussion started \cite{art:Lim2008, art:Lim2015} where an original $G_2$ solution was found.
Future investigations of this new solution will deal with a deeper analysis of the spike crossing phenomenon and its relationship to the structure formation problem in late time cosmology.

\subsection*{Discussion}

It is a general feature of solutions of partial differential equations that spikes occur~\cite{inbook:Wei2008}. 
Spikes occur in generic solutions of the EFE of general relativity~\cite{thesis:Lim2004}.  
Indeed, when a self-similar solution of the EFE is unstable, spikes can arise near such solutions.
Spikes were originally found in the context of vacuum orthogonally transitive (OT) $G_2$ models
~\cite{art:Limetal2009,art:BergerMoncrief1993,art:RendallWeaver2001,art:Lim2008}. 
Recently, Lim generalized an exact OT $G_2$ spike solution
to the non-OT $G_2$ case by applying Geroch's transformation 
on a Kasner seed~\cite{art:Lim2015}. 
Numerical
evidence has been presented~\cite{art:Limetal2009} that spikes in the Mixmaster 
regime of $G_2$ cosmologies are transient and recurring, 
supporting the conjecture that the generalized Mixmaster 
behavior is asymptotically non-local where spikes occur, leading to a violation
of BKL locality at exceptional (non-generic) spacetime points.

Scalar fields are ubiquitous in the early universe in modern theories of theoretical physics.
In the approach to the singularity the scalar field is dynamically massless \cite{book:Coley2003,art:CarrColey1999}.
Thus including massless scalar fields in early universe cosmology is important.
The EFE of a minimally coupled scalar field with a timelike gradient
are formally the same as those of an irrotational stiff perfect fluid.
BKL \cite{art:BK1981} studied perfect fluid models with the ``stiff matter" equation of
in the neighborhood of the
initial singularity where the Kasner and the Jacobs  relations are violated 
\cite{book:Coley2003,art:CarrColey1999}), in which case 
the collapse is generically 
described by a monotonic (but anisotropic) contraction of space along all
directions \cite{art:BK1981}. 

In \cite{art:ColeyLim2016} we studied spikes in the massless scalar field/stiff perfect fluid case.
In particular, we obtained a new class of exact non-OT $G_2$ stiff fluid spike solutions,
generalising the vacuum solutions of~\cite{art:Lim2015}, by 
applying the stiff fluid version of Geroch's transformation~\cite{art:Geroch1971,art:Geroch1972,art:GarfinkleGlassKrisch1997,art:Stephani1988} 
on a Jacobs seed. 
The dynamics of the stiff fluid spike solution is qualitatively different from 
that of the vacuum spike solution, in that the stiff fluid spike solution can end up with a permanent spike.

Due to
gravitational instability, the
inhomogeneous spikes leave small residual imprints on matter in the
form of matter perturbations in the early universe.
We investigated  the imprint of spikes on matter and structure formation in the
massless scalar field/stiff perfect fluid models in \cite{art:ColeyLim2012,art:LimColey2014}.

Both the incomplete spikes and the recurring spikes, related to self-similar solutions 
such as the Kasner solution and the FL model, may also occur at late times, and may also
cause spikes that might lead to further matter inhomogeneities, albeit non-generically, which
might lead to the existence of exceptional structures on large scales. 
Permanent spikes
in LTB models were studied in \cite{art:ColeyLim2014}.
In future work we shall consider as seeds
different stiff fluid models that do not isotropize or homogenize at late times 
but contain the FL model as a special (saddle) solution,
to investigate possible spike behaviour to the future.
Possible exact seed solutions include tilting
and non-tilting  stiff fluid Bianchi type $\text{VI}_h$, $\text{VI}_0$ and $\text{VII}_h$ models \cite[pp 359--360]{book:Exactsol2002}.  
These stiff fluid solutions are generated by Wainwright et al.\ \cite{art:Wainwrightetal1979} transformation on a vacuum seed; 
so it is also of interest to determine whether Wainwright et al.\ transformation and Geroch transformation commute.

\section*{Acknowledgments}

This work was supported, in part, by NSERC of Canada and AARMS.

\end{document}